# Optical Solver of Combinatorial Problems: Nano-Technological Approach

Eyal Cohen[*]     Shlomi Dolev[*]     Sergey Frenkel[†]
Boris Kryzhanovsky[‡]     Alexandr Palagushkin[‡]
Michael Rosenblit[§]     Victor Zakharov[†]

October 29, 2018

## Abstract

We report the first steps in creating an optical computing system. This system may solve NP-Hard problems by utilizing a setup of exponential sized masks. This is exponential space complexity but the production of those masks is done with a polynomial time preprocessing. These masks are later used to solve the problem in polynomial time. We propose to reduced the size of the masks to nano-scaled density. Simulations were done to choose a proper design, and actual implementations show the feasibility of such a system.

## 1 Introduction

It has been shown that by copying a large amount of data at once, it is possible to create a preprocessed data structure of exponential size in a polynomial time complexity. Previous work [1] suggested using this method in creating exponential sized masks. These masks are used to create devices that may solve NP-Complete problems. By choosing a subset of the masks we can check a certain instance of the problem.

[*]Department of Computer Science, Ben Gurion University of the Negev, 653 Be'er Sheva 84105, Israel
  [†]Institute of Informatics Problems, Russian Academy of Science, Russia
  [‡]Scientific Research Institute for System Studies (NIISI), Russian Academy of Sciences, Russia
  [§]Ilze Katz Institute for Nanoscale Science and Technology, Ben Gurion University of the Negev, 653 Be'er Sheva 84105, Israel

We suggested building such exponential sized masks in nano-scaled density [2]. When each pixel of the masks would be microscopic, it will allow us to build smaller masks which hold more information.

These masks would be used in an optical setup. Each path or pixel of these masks is actually performing a different calculation. Those calculations are performed optically so it is easier to perform parallel calculations. The first step in creating a device like this would be to test its optical properties. By building layers of the masks one over the other we can test the transmittance of the setup.

Creation of the masks would be done by the preprocessing algorithm presented in our earlier work [2]. This algorithm uses the copying technique to create exponential sized masks in polynomial time. These masks will be layered one over another and their overall setup will determine the transmittance of the system. This transmittance would give us the result of a computation.

A simulation of such systems were done using programs that simulate light propagation, by solving the Maxwell equations. Simulations were done to analyze the different structures, allowing us to choose a proper structures for the masks. Simulations were testing the ability to transmit or block light without unwanted effects like diffraction, reflection, refraction, or crosstalk between channels.

After a general structure was chosen, physical test setups were created. First only one layer of single masks were created, agreeing with the preprocessing algorithm and the needs of the system. Later, layers of masks were built as suggested. Different set of masks were used creating setups of up to 4 layers.

## 2  Optical Solutions for Hard Problems

Optical computers are devices that use light rather than electricity in order to preform calculations. These devices benefit from the advantages of light over electricity. Light beams for example may intersect without interference, unlike electric currents.

Integrating a processing unit with fiber optics and laser on silicon may improve a computer's speed and lower energy loss. This may be due to elimination of resistance, crosstalks, soft errors, and the like which appear when electronic connections are used.

One approach in creating an optical processing unit would implement optical gates. These would be used to map the current VLSI designs to general-purpose all-optical processors [3, 4]. However, the overhead involved in creating this general-purpose all-optical computer would be too high in comparison to a processor designed to perform specific primitives. Special purpose optical devices designed to solve a limited set of hard problems (e.g., NP- and P♯ complete or EXP-TIME complete) can be utilized by the industry right away. Researching these devices can later lead to and help building general purpose optical computers.

Many researchers aim at finding efficient solutions to hard combinatorial problems. Currently only exhaustive searches guarantee finding the optimal solution [5]. For large problems, these methods result in exponential execution time.

Optical computers may be different from their electronic counterparts just as electronic computers are not structured like mechanical computers [6]. Many features of light could be exploited to ease computation. For example, some success in using many beams in free space for computing has been recently reported [7]. These designs are based on parallel optical multiplication. The use of similar multiplication devices to solve NP-complete problems was also suggested [8, 9].

Reif et al. [10] prove that the problem of ray tracing, in three-dimensional optical systems consisting of a finite set of mirrors with endpoints that are rational coordinates, is PSPACE-hard. In [11] we suggest using a mapping similar to the non-deterministic Turing machine by (amplifying and) splitting beams. The mapping can be viewed as a theoretical proof of the existence of a solution rather than an efficient solution. Knowing, that a solution exists for a (bounded) NP-complete problem instance, we seek the most efficient solution in terms of: the number of beams used; the number of optical elements (or location in space used to represent a computation state); the energy needed; the maximum number of beams that should be split from a single source beam (fanout); and the number of locations a beam needs to visit (and possibly split) from its creation until its final detection on arrival.

We are not solving NP-complete problems with polynomial (in the input length) space. We suggest that an optical approach may be promising in solving larger instances of hard combinatorial problems. In fact, we use exponential space to solve exhaustively an instance in linear time.

## 2.1 Optical Solutions for Hamiltonian-cycle and Permanent Problems

Different designs for an optical architecture, that solves the NP-complete Hamiltonian-cycle problem, have been suggested. The Hamiltonian-cycle problem is one of the most interesting problems solved using optical computing. The architecture we have suggested in [11] and the architectures suggested in [12, 13] are based on splitting and delaying the light beams. The Hamiltonian-cycle problem can also be solved following either one of the two approaches of the traveling beams architecture we have suggested in [1]. The first approach in [1] can also be used to solve the permanent problem. The first approach involves mapping the graph nodes to physical locations in space and the propagation of beams according to the edges of the input graph instance. The second approach propagates beams along a computation tree such that the leaves represent all possible solutions and the delay in propagation from the root to each leaf corresponds to the value of the specific combination. We also present in [1] a very different architecture called the coordinated holes in a mask-made black box (see Figure 1). In this architecture a set of masks with holes is chosen from $n(n-1)$ pre-computed masks, according to the input instance. A solution to the problem instance exists only if the combined masks do not block all beams.

Optical copying is an important and powerful computation technique. Many examples could be found, e.g. [14]. In [15] we implement a specific set of primitives that allows us to have an efficient pre-processing stage based on mask creation by copying matrices or matrices columns. In addition, in [15, 16] we present a technique for solving problems that are hard-in-average, namely, the shortest vector problem and integer version of the permanent. We will describe the solution for the binary permanent below.

## 2.2 Hamiltonian-Cycle

The Hamiltonian-cycle problem tries to find a closed path that goes through all vertices of a given graph, while visiting each vertex exactly once. We solve the directional Hamiltonian cycle problem where edges of the graph may be directional. The Hamiltonian-cycle problem is NP-complete problem and as such there is no known polynomial time algorithm to solve it. We use exhaustive search, which imply that the number of possible cycles grows factorialy with the number of vertices of the graph.

A full graph with $n$ vertices could be represented by a binary matrix. The matrix has $n(n-1)$ and $(n-1)!$ rows, where each column represents an edge in the full graph, and each row represents a possible path. Every matrix element identified by a possible path and an edge. Examples of $n = 3$ and $n = 4$ matrices are $M_3$ and $M_4$ respectively,

$$M_3 = \begin{pmatrix} 1 & 0 & 0 & 1 & 1 & 0 \\ 0 & 1 & 1 & 0 & 0 & 1 \end{pmatrix}$$

$$M_4 = \begin{pmatrix} 1 & 0 & 0 & 0 & 1 & 0 & 0 & 0 & 1 & 1 & 0 & 0 \\ 1 & 0 & 0 & 0 & 0 & 1 & 1 & 0 & 0 & 0 & 0 & 1 \\ 0 & 1 & 0 & 0 & 0 & 1 & 0 & 1 & 0 & 1 & 0 & 0 \\ 0 & 1 & 0 & 1 & 0 & 0 & 0 & 0 & 1 & 0 & 1 & 0 \\ 0 & 0 & 1 & 1 & 0 & 0 & 0 & 1 & 0 & 0 & 0 & 1 \\ 0 & 0 & 1 & 0 & 1 & 0 & 1 & 0 & 0 & 0 & 1 & 0 \end{pmatrix}.$$

A processing system should receive a vector as an input which represents the existence of edges in the graph. If an edge exist on the input graph, the corresponding vector element has the value 0 and if it does not exist its value is 1. The edges are organized in vectors in the appropriate matrices. The organization of vectors that correspond to matrices $M_3$ and $M_4$ are represented by vectors $V_3$ and $V_4$ respectively.

$$V_3 = \begin{pmatrix} e_{12} \\ e_{13} \\ e_{21} \\ e_{23} \\ e_{31} \\ e_{32} \end{pmatrix}$$

$$V_4 = \begin{pmatrix} e_{12} \\ e_{13} \\ e_{14} \\ e_{21} \\ e_{23} \\ e_{24} \\ e_{31} \\ e_{32} \\ e_{34} \\ e_{41} \\ e_{42} \\ e_{43} \end{pmatrix}.$$

The system should multiply the vector by the matrix. If the result vector contains a 0 (a zero) the Hamiltonian cycle is possible.

The algorithm presented in [1] is an extending algorithm. That means that the algorithm uses Mn to create $M_{n+1}$. It extends $(n-1)! \times [n(n-1)]$ matrix to a $n! \times [n(n+1)]$ matrix. The algorithm uses an optical copying method to produce the masks. We use optical lithography to copy entire sections of the original matrix to the new matrix. Each iteration the algorithm copies an entire column of the original matrix. Thus $(n-1)!$ matrix elements are copied at once. When this is done $n$ times, we get a column with $n!$ matrix elements. This eliminates the need of factorial in the time complexity, even though the output is a factorial sized matrix.

## 2.3 The Permanent Problem

The Permanent problem is defined by

$$Perm(A) = \sum_\sigma \prod_{i=1}^n A_{i\sigma(i)},$$

where $A$ is an $nn$ matrix, and the summation is over all permutations $\sigma$ on $n$ elements. An optical processing system can solve the binary matrix permanent problem. A $2 \times 2$ binary matrix could be rearranged to be represented by a vector by reorganizing the elements and taking their logical inverse,

$$A_2 = \begin{pmatrix} a_{11} & a_{12} \\ a_{21} & a_{22} \end{pmatrix}$$

$$V_2 = \begin{pmatrix} \neg a_{11} \\ \neg a_{12} \\ \neg a_{21} \\ \neg a_{22} \end{pmatrix}.$$

Multiplying the vector by an appropriate matrix could give us the multiplications we need to sum upon by using De Morgans laws,

$$M_2 = \begin{pmatrix} 1 & 0 & 0 & 1 \\ 0 & 1 & 1 & 0 \end{pmatrix}$$

$$M_2 \cdot V_2 = \begin{pmatrix} \neg a_{11} + \neg a_{22} \\ \neg a_{12} + \neg a_{21} \end{pmatrix} = \begin{pmatrix} \neg(a_{11} \cdot a_{22}) \\ \neg(a_{12} \cdot a_{21}) \end{pmatrix}.$$

Again, it shows that this problem is based on a vector multiplication by a matrix. This means that in order to calculate these multiplications, we can use the same architecture as the Hamiltonian-cycle. The sum of the multiplications could be done by collecting all beams (by lens for example), and using a detector that measures the intensity which determines the sum of the beams that managed to pass through the masks.

## 3 Design

The implementation of nano-scale optical computing devices will serve a proof of concept and in time may lead to advances in computing capabilities. We use lithography method to produce masks [17, 18]. The size of the mask will be in the range of $20 \times 20 cm^2$ to $10 \times 10 cm^2$ to solve an instance of the Hamiltonian-cycle when the number of nodes in the graph is $n = 15$. The masks will be produced with high precision and designed for wavelengths that will allow us to increase a density of the optical information pixels. The material that will be chosen will be transparent in the short wavelength range.

The resulting mask will be a monolithic design, meaning that light will propagate through a material with a high refraction index. Light will dissipate elsewhere where the refraction index is low. This is similar to light propagating through fiber optics.

We consider and test the methods of creating the masks. The first method considers a straight forward approach where masks block light exactly as the algorithm suggests.

A layer selection mechanism should be chosen. This mechanism will make it possible to activate any subset of the layers. There are several possibilities to select mask setup corresponds to problem under investigation. A selection mechanism can be based on MOEMS (MEMS), that will make it possible to activate any subset of the masks.

## 3.1 Masks Production

A straight forward approach is taken, where masks block light exactly as the algorithm suggests. An optical algorithms makes use of (exponentially large) masks to solve a number of NP-hard problems. High-resolution optical films are employed in the input and output planes of the optical system in order to synthesize a large binary matrix [1]. The influence of background noise, low light intensity, as well unwanted duplications was examined. We note that mask creation by an iterative optical synthesis process is an essential part of the optical computer design. The size of the mask, the manufacturing precision, and the possibility to increase the density of the optical information pixels are very important parameters for practical use. In order to make it so we use modern nanotechnologies implemented to create metallized photo-patterns. This includes the following principal directions:

(a) To develop a method of topologically transforming a computer generated masks to standard vector graphic format compatible with Laser writer and e-line. The working field will be divided into several sub-fields.

(b) To create a set of patterns using topological files of sub-fields and alignment marks to combine these sub-field patterns.

(c) Assembling of patterns into a working optical mask will be performed by projective lithography using AER photo multiplication with a decreasing factor of 10.

After the fabrication process we will get a set of nano-scale structures which will represent optical algorithms through masks implementing in set of samples each of which maps the graph nodes to the physical locations. In the final stage we will add a layer selection mechanism and enclose the masks in a miniature package.

# 4 Nano Based Masks Performance Simulation

When light propagates through the masks setup, eventually some of the light may pass through all the masks in some of the pixels. We call each of the paths a channel. We simulate this system of masks simulating monolithic nano-structures. Propagating light in nano scale monolithic structure behaves like a light inside a waveguide. The entire setup is attached to a light source, ensuring that every channel will get the same light level. There is a need for a careful design of the masks in order to avoid unwanted mode coupling, multimode-interferences, PDL, etc. That may result in errors in calculations when light from one channel arrives to another. We refer to this problem as channel crosstalk.

We simulated the architecture using the BeamProp (RSoft) and OptiFDTD (Optiwave) software.

We examined several structures, materials and wavelengths for our design to allow optimization of optical mask size. Design, simulation and tolerance analysis were done for monolithic structures based on silicon, taking into account fabrication ability, and possible photonics complication (mode coupling, input mode mismatch, polarization dependence, etc.). On the next stage we will implement fabrication nanotechnologies to create a set of monolithic patterns based on silicon wafers.

Simulations were done for silicon - silicon nitride structures for $n = 3, 4 \ldots$ masks. Simulations for silicon - silica should be the next step.

## 4.1 Simulation Model Setup

A preliminary simulation was done to choose an optimal set of variables such as the width of the pixel, the difference in the width of two neighboring pixel, the distance between the channels, and the wavelength. The simulation was done for a $3 \times 3$ setup, where only the central pixel should allow light to propagate. We can change the width of the neighboring pixel and compare the amount of light emitted through the central pixel to the light emitted through other pixels. We would like to maximize the ratio of the emission of the central pixel over the other pixels. The values of the actual simulation were selected as the optimal values by different aspects of the photonics complications.

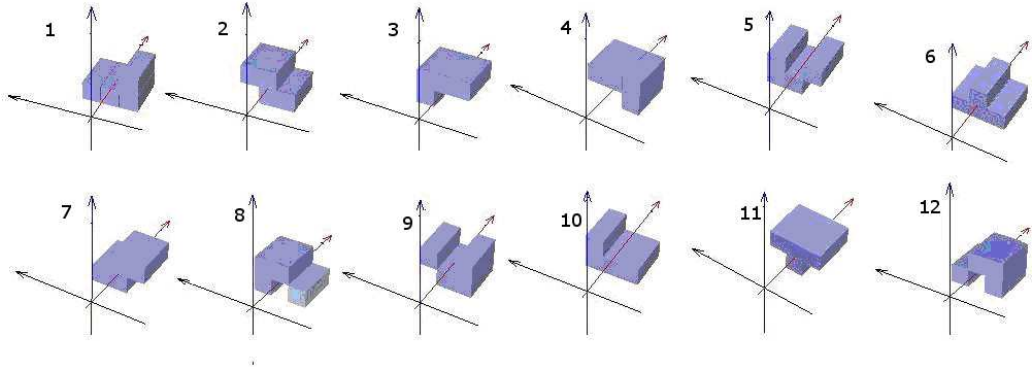

Figure 1: Example masks for a Hamiltonian cycle with $n = 4$

The setup we chose is $n = 4$ Hamiltonian cycle problem (Figure 1). In all the masks four pixels should allow light to propagate. The desired result depends on the mask setup configuration and show how many pixels can emit light.

All simulations of this example were done for silicon-air masks and a waveguide that directs the light into the setup. The free space wavelength used was $\lambda = 1.3\mu m$. The refraction index $n = 3.5$ used for inside the material is suitable for silicon. The distance between the center of neighboring channel is $0.6\mu m$.

Figure 2 shows a single mask #2 (Figure 1) with different polarization of the input light. The two output patterns differ and would suggest that in this specific mask pattern, a Transverse Electric (TE) polarization is preferable as its transmission is more similar to the shape of the mask.

Next we have to check if we avoid crosstalk, that is, making sure that light that propagates in one channel continues to do so, and do not stimulate propagation in another channel. Several mask setups were simulated to check if it satisfies those demands. We would like to check what happens in the system when only one pixel should allow propagation.

Figure 3 presents the output of a setup of three masks. The masks were $1, 2, 4$ (Figure 1). Using the setup in Figure 3 allows light to propagate only through the lower right pixel. The light on the other pixels should dissipate or scatter.

In Figure 4, another mask was added such that the masks used are $1, 2, 4, 10$ (Figure 1). This setup is checking if adding more masks keeps the light propagating through the lower right pixel, and that it does not stimulate propagation on other pixels.

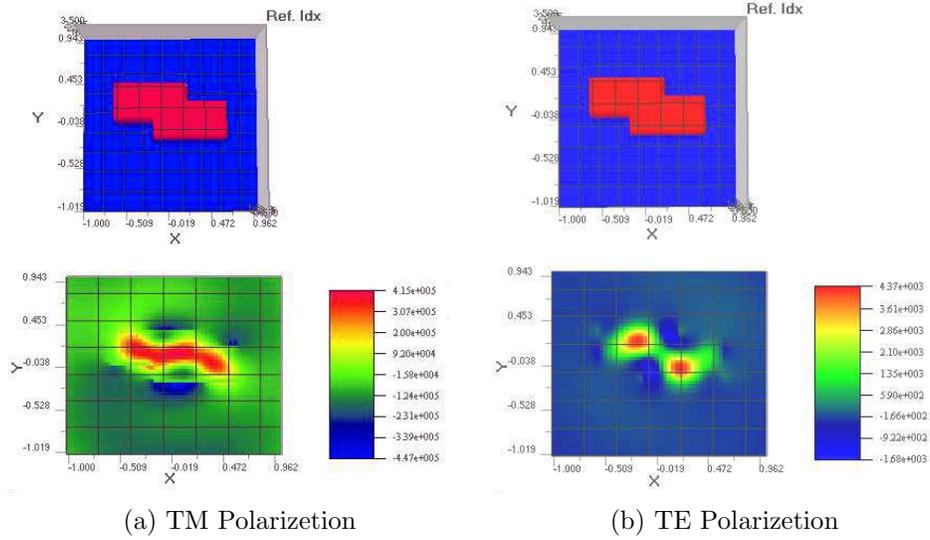

(a) TM Polarizetion      (b) TE Polarizetion

Figure 2: Transmittance results of one mask (drawn above) for (a) TM and (b) TE polarizations

Another important setup simulation was done using masks that should block all the light. This time we used masks $1, 2, 3$ and the output is displayed in Figure 5. The same results should be achieved using all masks since this setup blocks all the pixels and more masks will only lower the transmittance.

## 5 Masks Implementation

The next step after the simulations are actual nanoscaled masks productions. We would like to test a system with few layers of masks layered one over the other. This experiment will be created for testing purposes and will not include a layer selection mechanism. Thus we would like to test many configurations of different static layers. The experiment should agree with the simulations and produce the desired output as the theoretic algorithm suggested.

The model chosen for the production is the Hamiltonian cycle problem with $n = 5$. As described in the previous sections, each mask of this problem should have $4! = 24$ pixels. To ease the process of production and pre-production we decided to arrange these pixels in squares. The width and height of the square should be $\left\lceil \sqrt{(n-1)!} \right\rceil$. Thus the pixels are arranged in a square of $5 \times 5$ where the last pixel won't be used. This is chosen instead of the $6 \times 4$ pixel orientation.

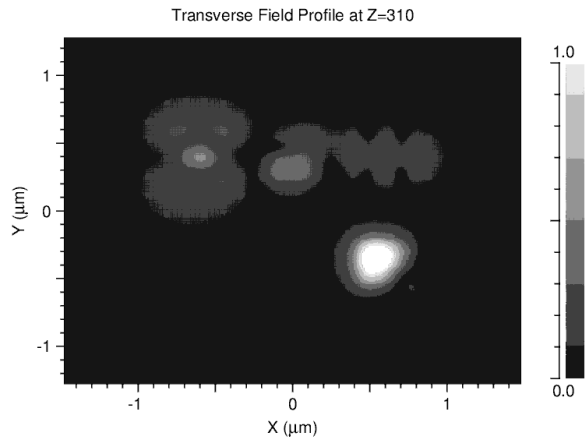

Figure 3: Output of a 3 masks setup. Masks $1, 2, 4$ from Figure 1

Layers of masks representing the Hamiltonian cycle problem with $n = 5$ would always transmit light when the number of mask is lower than 4. It will take at least 4 masks in order to build a layered setup that will not transmit light. For example the masks instance $\{e_{12}, e_{13}, e_{14}, e_{15}\}$, include all paths going out of vertex $v_1$. This instance should block all the light since there is no possible Hamiltonian cycle.

We chose to limit the number of layers in the experiment to 4, as the lowest number of masks that yield layers setups that would block all the light. We have 20 different masks available to choose from, and we want to try each possible combination of those masks. This leads to 4845 possible combinations of the different masks. These different combinations could be treated as different instances of the problem. These combinations would fit on a square of $70 \times 70$ where each row holds 70 different instances. On the last row only the first 15 of the instances are used, while 55 are unused.

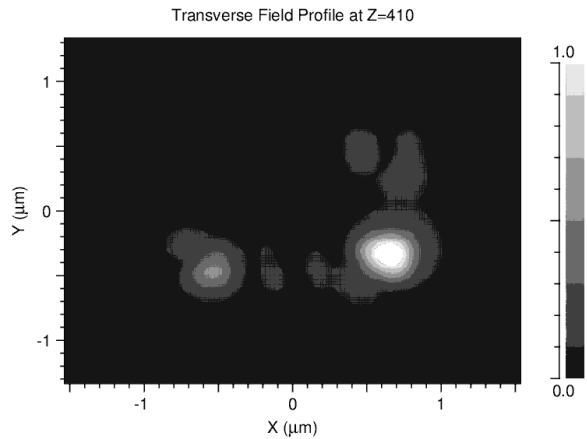

Figure 4: Output of a 4 masks setup. Masks $1, 2, 4, 10$ from Figure 1

The size of each pixel on the mask was chosen to be $1.5 \times 1.5 \mu m^2$ and the distance between two pixels was also $1.5 \mu m$. This in turn sets the size of the masks to be $15 \times 15 \mu m^2$. The distance between two masks for different instances, was also chosen to be $15 \mu m$. This would yield an array of instances of $2.1 \times 2.1 mm^2$. Figure 6 shows the design of the entire array, and a zoom in on one mask inside the array, composed of $5 \times 5$ pixels.

## 5.1 Mask Implementation Issues

Currently, a real mask topology has been developed. Some technological and physical questions of such topology practical implementation on the base of E-beam equipment have been investigated in NIISI, Moscow Russia. A specific attribute of this topology is very high density of the pixels and pseudorandom in their location. Therefore, it is difficult to use a direct photocomposition. Instead, a decomposition (splitting) of the original topological array into several photolithographic masks and their sequential assembling has been used. The masks are (implemented) on the plates covered by chrome and of a size $4 \times 4$" ($102 \times 102 mm^2$) of thick $2.4 \pm 0.2 mm$ with working field $80 \times 80 mm^2$.

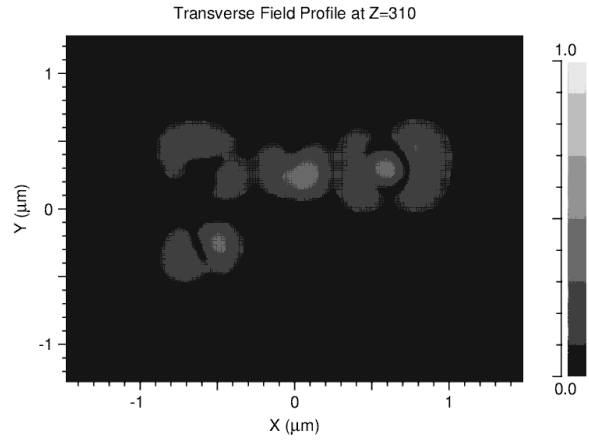

Figure 5: Output of a 3 masks setup. Masks $1, 2, 3$ from Figure 1

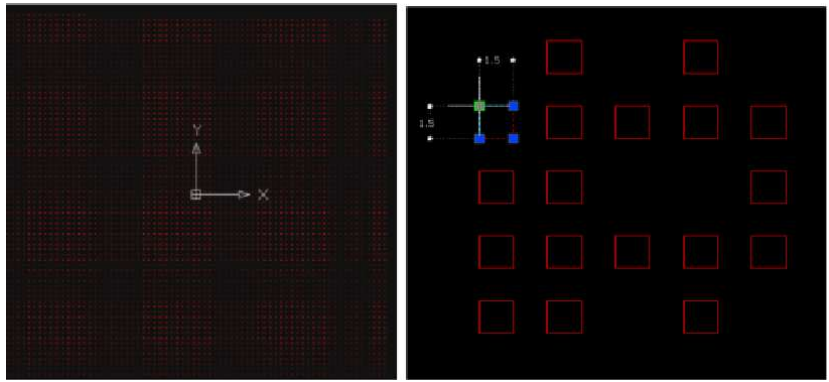

Figure 6: The design of the array of masks (Left), and an enlargement of one mask (Right).

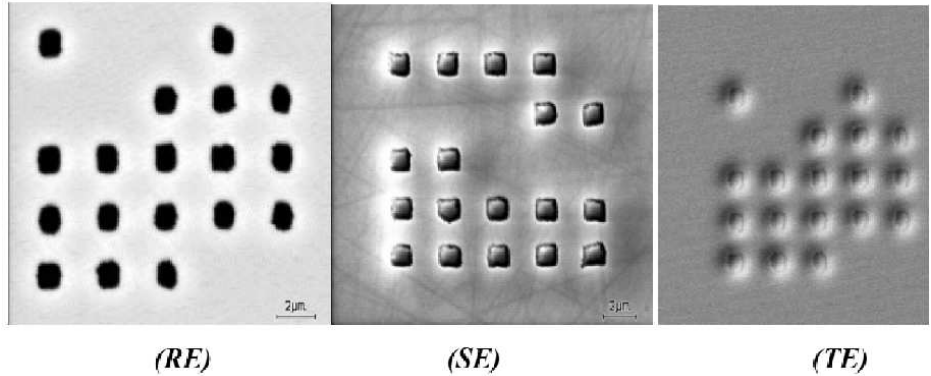

Figure 7: Images of different masks using electron microscope. Obtained by reflected electrons(RE), secondary electrons (SE) and in the topographic contrast (TE).

Assembling of the separate fragments of the mask were implemented with Carl Zeiss photosystem AER-topology (secondary emission photocell, or photomultiplier) with 10-time size reducing.

Electro-lithographic mask manufacturing technique were developed and-tuned for Electron Beam scanning lithographic system ZMR 20 (Carl Zeiss, Germany) with an additional graphic generator controlled by the NanoMaker software.

The masks were investigated by optical microscope NU-2E with a program of a TV based system of image analysis Image-Pro Plus (Media Cybernetics,USA).

Figure 7 shows different electronic images (photography) of different $5 \times 5$ masks with pixel groups obtained by reflected electrons(RE), secondary electrons (SE) and in the topographic contrast (TE).

The passing of the light through the mask was measured for the white halogen lamp by the condenser microscope system NU-2E (Carl Zeiss) with a digital TV camera. The image obtained was proceeded by the program Image-Pro Plus (Media Cybernetics).

The light distributions after the mask passing were studied. Figure 8

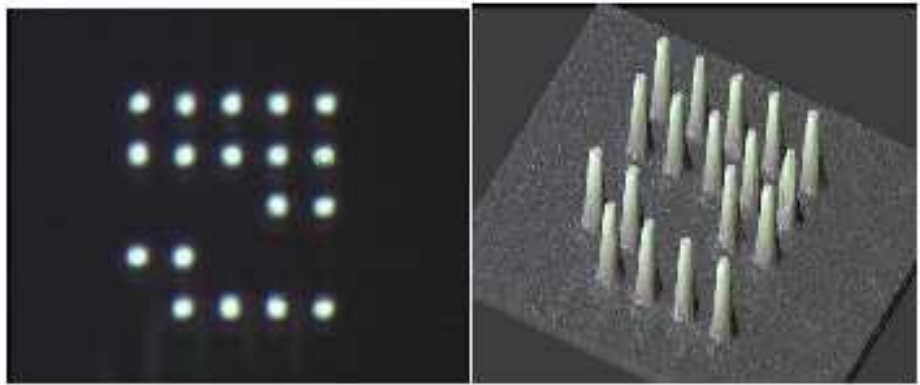

Figure 8: Distribution of the light passing through the mask holes

shows the optical photography of the pixel groups. It is important that both the distributions for white light and for the RGB spectrum are equal practically for all pixels of the groups.

# 6    Concluding Remarks

We presented a system which deals with combinatorial tasks, which is useful especially when there are real-time constraints. Theory and algorithms were created and tested for a micro optical architecture for solving instances of NP-Hard problems using Nanotechnology. Simulations and the productions of several masks in our laboratory demonstrate the feasibility of the architecture. We are encouraged to believe that our new designs will be used in practice.

## 6.1    Future Plans

We plan to extend the production of the masks to multilayered setups that allows us to further test the architecture. More simulations should be done for different monolithic designs. A layer selection mechanism that will allow to activate any subset of the layers should be chosen. This mechanism should be based on MOEMS, tunable plasmonics or other mechanisms.

# 7 Acknowledgements

This research was supported by a grant from the Ministry of Science & Technology, Israel & the Russian Foundation for Basic Research, the Russian Federation.